\documentclass[aps,prl,twocolumn,superscriptaddress]{revtex4}
\usepackage{graphicx}
\usepackage{latexsym}
\usepackage{amssymb}
\usepackage{amsmath}
\usepackage{amsfonts}
\usepackage{bm}
\usepackage{multirow}
\usepackage{color}

\newcommand{\dd}{\mathrm{d}}

\newcommand{\dsZ}{\mathbb{Z}}
\newcommand{\Tr}{\mathop{\mathrm{Tr}}}

\newcommand{\vect}[1]{{\bm{#1}}}

\newcommand{\beq}{\begin{equation}}
\newcommand{\eeq}{\end{equation}}
\newcommand{\beqn}{\begin{eqnarray}}
\newcommand{\eeqn}{\end{eqnarray}}

\begin{document}

\title{Self-dual Quantum Electrodynamics on the boundary of $4d$ Bosonic
\\ Symmetry Protected Topological States}




\author{Zhen Bi}

\author{Kevin Slagle}

\author{Cenke Xu}

\affiliation{Department of physics, University of California,
Santa Barbara, CA 93106, USA}

\begin{abstract}

We study $3d$ (or $(3+1)d$) Quantum Electrodynamics (QED) realized
on the boundary of $4d$ (or $(4+1)d$) bosonic symmetry protected
topological (BSPT) states, using a systematic nonlinear sigma
model (NLSM) field theory description of BSPT states developed in
Ref.~\onlinecite{xuclass}. We demonstrate that many of these QED
states have an exact electric-magnetic duality due to the symmetry
of the BSPT states in the $4d$ bulk. The gauge charge and Dirac
monopole both carry projective representations of the bulk
symmetry, and the emergent gapless photons of the QED phase also
transform nontrivially under the bulk symmetry. Some of these QED
boundary states can be further driven into a $3d$ $\mathbb{Z}_2$
topological order, and the statistics and symmetry transformation
of its point particle and vison loop excitations guarantee that
this topological order cannot be driven into a trivial confined or
Higgs phase. With a finite fourth dimension, the entire system
becomes a $3d$ lattice, the self-dual QED and the $\mathbb{Z}_2$
topological order can coexist on two opposite boundaries
respectively, which together constitute an exotic $3d$ self-dual
``topological photon phase".

\end{abstract}

\pacs{}

\maketitle


\section{1. Introduction}

Symmetry protected topological (SPT) phases, a new type of quantum
disordered phase pioneered in Ref.~\onlinecite{wenspt,wenspt2},
are intrinsically different from trivial direct product state,
when and only when the system has certain symmetry $G$. In terms
of its phenomena, a SPT phase on a $d-$dimensional lattice should
satisfy at least the following three criteria:

({\it i}). On a $d-$dimensional lattice without boundary, this
phase is fully gapped, and nondegenerate;

({\it ii}). On a $d-$dimensional lattice with a
$(d-1)-$dimensional boundary, if the Hamiltonian of the entire
system (including both bulk and boundary Hamiltonian) preserves
symmetry $G$, then this phase is either gapless or gapped but
degenerate.

({\it iii}). The boundary state of this $d-$dimensional system
cannot be realized as a $(d-1)$-dimensional lattice system with
the same symmetry $G$.

Notice that the second criterion ({\it ii}) implies the following
two possibilities: On a lattice with a boundary, the system is
either gapless, or gapped but degenerate. For a $1d$ SPT state,
its $0d$ boundary must be degenerate, which forms a projective
representation of the symmetry group; for a $2d$ SPT state, its
$1d$ boundary can be either gapless, or degenerate due to
spontaneous discrete symmetry breaking; for a $3d$ SPT state, the
degeneracy of its $2d$ boundary can correspond to either
spontaneous breaking of $G$, or correspond to certain topological
degeneracy at the boundary. Which case occurs in the system will
depend on the detailed Hamiltonian at the boundary of the system.
For example, with strong interaction, the boundary of a 3d
topological band insulator can be driven into a nontrivial
topological
phase~\cite{TI_fidkowski1,TI_fidkowski2,TI_qi,TI_senthil,TI_max}.
And one natural candidate $2d$ boundary state of most $3d$ bosonic
SPT states is a $2d$ $Z_2$ topological order with $e$ and $m$
excitations both carrying fractional degrees of
freedom~\cite{xuclass,senthilashvin}.

In this paper we will investigate the $3d$ boundary states of $4d$
bosonic SPT states. All the possible boundary states discussed
above can occur in our case, but there is one new possibility
which does not occur in lower dimensions: the $3d$ boundary can be
a deconfined gapless photon state, which does not exist in lower
dimensions with gapped matter fields due to the well-known fact
that the $(2+1)d$ compact QED with gapped matter fields is always
confined due to the proliferation of Dirac monopole in the
space-time. However in $(3+1)d$, a compact QED can have a deconfined
photon phase with gapless photon excitations, and deconfined
electric charge (denoted as $e$) and Dirac monopole
(denoted as $m$) excitations. This boundary gapless photon must be
very unusual, because based on the definition of an SPT state, this
boundary state cannot be realized in $3d$ without the bulk or the
opposite boundary.

Please note that this photon phase only exists at the $3d$
boundary; namely the bulk is still fully gapped and nondegenerate.
Just like the $2d$ topological order at the boundary of a $3d$ SPT
state, the $e$ and $m$ excitations of the photon phase must carry
a nontrivial representation (or projective representation) of the
symmetry groups, which implies that the system cannot be driven into a
trivial confined or Higgs phase with a gapped and nondegenerate
ground state by condensing $e$ or $m$ excitations. The quantum
number of $e$ and $m$ excitations can be computed systematically
using the NLSM field theory developed in
Ref.~\onlinecite{xuclass}. Besides the quantum numbers carried by
$e$ and $m$, we are going to show that in many cases the boundary
photon phase has an exact ``self-dual" symmetry. In this work we
are going to describe two examples in detail. In the first
example, the self-dual symmetry is the physical time-reversal
symmetry, and in the second example it is a $Z_4$ symmetry.

A $4d$ system with infinite size is unrealistic. However, our
study of the $3d$ boundary of a $4d$ SPT state can lead to exotic
phases in $3d$ as well. We can imagine making a thin slab of the
$4d$ SPT state, namely a $4d$ system with a finite fourth
dimension. Then the entire system becomes three dimensional, but
one can realize two different $3d$ states on the two opposite
boundaries. For the first example state in which time-reversal
plays the role of a self-dual symmetry, we can realize the $3d$
self-dual photon phase on the top surface, but realize a fully
gapped topological order on the bottom surface. Then at low energy
only the photon phase on the top surface becomes visible, which by
definition is a phase that cannot be realized in $3d$ at all. Only
at higher energy will the topological order on the bottom surface
be exposed. By contrast, for the example where the $Z_4$ symmetry
plays the role as self-duality, it seems this photon phase cannot
be driven into any gapped topological order without breaking the
$Z_4$ symmetry.

We note that in Ref.~\onlinecite{mcgreevy2013,mcgreevy2014}, a QED
state on the $3d$ boundary of a $4d$ bosonic short range entangled
(BSRE) state was also studied, and this QED state can have a
maximum $SL(2, \mathbb{Z})$ duality symmetry. But in that case the
$4d$ BSRE state does not need any symmetry to be nontrivial
because the $3d$ boundary QED state is an ``all fermion state";
namely its $e$ and $m$ excitations are both
fermions~\cite{wangpottersenthil}, which is a fact robust against
any symmetry breaking. This bulk BSRE state is also an
``invertible topological state" discussed in
Ref.~\onlinecite{kongwen,wenso,kapustin1,kapustin4,freed2014}.
However, in our case, the $4d$ system is a nontrivial SPT state
when and only when the system has certain symmetry. When the
symmetry is broken, the system becomes a trivial bose Mott
insulator.



\section{2. Self-dual photon phase with $Z_2 \times \mathcal{T}$ symmetry}

Let us start with a $4d$ SPT state with $Z_2$ and time-reversal
symmetry ($\mathcal{T}$) symmetry only. This state can be
described by the following NLSM field theory with a six component
unit vector $\vect{n}$: \beqn \mathcal{S} &=& \int d^4x d\tau \
\frac{1}{g} (\partial_\mu \vect{n})^2 \cr\cr &+&
\frac{i2\pi}{\Omega_5} \epsilon_{abcdef} n^a \partial_{x} n^b
\partial_{y} n^c \partial_{z} n^d \partial_{x_4} n^e \partial_\tau n^f.
\label{o6}\eeqn Here $|\vect{n}| = 1$, and $\Omega_5$ is the
volume of the five dimensional unit sphere. In Eq.~\ref{o6}, when
the coupling constant $g$ is larger than some critical value, the
system is in a quantum disordered phase with a fully gapped and
nondegenerate bulk spectrum, and according to
Ref.~\onlinecite{xuclass}, different SPT states correspond to
different symmetry transformations on $\vect{n}$ that keep the
entire action, including the topological $\Theta-$term invariant.
In this work we primarily consider the state that corresponds to
the following transformation as an example: \beqn Z_2 &:& \vect{n}
\rightarrow - \vect{n}, \cr\cr \mathcal{T} &:& n_1
\rightleftarrows n_4, \ \ n_2 \rightleftarrows - n_5, \ \ n_3
\rightleftarrows n_6. \label{sym} \eeqn In fact, even without the
$\mathcal{T}$ symmetry this state is already a nontrivial SPT
state, and this state is just a $4d$ generalization of the $2d$
SPT with $Z_2$ symmetry~\cite{levingu}, which is described by a
$(2+1)d$ NLSM with a four component unit vector~\cite{xusenthil}.
Also, this state can be viewed as a $4d$ bosonic integer quantum
Hall state with U(1) symmetry broken down to $Z_2$ (see more
details in appendix {\bf B}).

In this work we will focus on the $Z_2 \times \mathcal{T}$
symmetry. But, Eq.~\ref{o6} actually can also describe SPT states
with much larger symmetries. Let us parameterize the six-component
vector $\vect{n}$ as $\vect{n} = (\cos\alpha \vect{N}, \sin\alpha
\vect{M})$, where $\vect{N}\sim (n_1, n_2, n_3)$ and $\vect{M}
\sim (n_4, n_5, n_6)$ are both three-component unit vectors. We
also tentatively introduce two more SO(3) symmetries to the system
with $\vect{N} $ and $\vect{M}$ transforming as vectors under the
two SO(3) symmetries respectively, although these exact SO(3)
symmetries are unimportant to the main physics we are going to
discuss. (Introducing extra symmetries and eventually breaking
them has proved to be a very helpful trick for field theory
analysis, as shwon in Ref.~\onlinecite{senthilashvin}.) At the
$(3+1)d$ boundary, Eq.~\ref{o6} will reduce to a $(3+1)d$ NLSM
with a Wess-Zumino-Witten (WZW) term~\cite{xuclass} at level-1:
\beqn \mathcal{S} &=& \int d^3x d\tau \ \frac{1}{g} (\partial_\mu
\vect{n})^2 \cr\cr &+& \int_0^1 du \frac{i2\pi}{\Omega_5}
\epsilon_{abcdef} n^a \partial_{x} n^b
\partial_{y} n^c \partial_{z} n^d \partial_{\tau} n^e \partial_u
n^f. \label{o6wzw} \eeqn Just like all WZW terms, the last
term in Eq.~\ref{o6wzw} is equal to the volume of the target space
$S^5$ enclosed by the trajectory of $\vect{n}$ under a periodic
evolution. $u \in [0,1]$ is an extra parameter introduced and
$\vect{n}(\vect{x}, \tau, u = 0) = (0,0,0,0,0,1)$ while
$\vect{n}(\vect{x}, \tau, u = 1) = \vect{n}(\vect{x},\tau)$.

The physical meaning of this WZW term becomes explicit when we
reduce this WZW term on a hedgehog monopole of $\vect{N} \sim
(n_1, n_2, n_3)$, which is a point defect ($\vect{N}$ is
normalized to be a three-component unit vector). Since the
hedgehog monopole is a singularity of $\vect{N}$, then at the
hedgehog monopole the six component unit vector $\vect{n}$ will
reduce to another three-component unit vector $\vect{M} \sim (n_4,
n_5, n_6)$, and this WZW term reduces to a $(0+1)d$ WZW model for
$\vect{M}$ (for more details please see appendix {\bf C}): \beqn
\mathcal{S}_{hm} = \int d\tau \ \frac{1}{g^\prime} (\partial_\tau
\vect{M})^2 + \int_0^1 du \frac{i2\pi}{\Omega_2} \epsilon_{abc}
M^a \partial_\tau M^b \partial_u M^c. \label{0dwzw}\eeqn With the
extra SO(3) symmetries, the ground state of this $(0+1)d$ field
theory Eq.~\ref{0dwzw} is two fold degenerate: \beqn | m \rangle =
\left( \cos(\theta/2) e^{i\phi/2}, \ \sin(\theta/2) e^{-i\phi/2}
\right)^t. \eeqn Here we have used the standard parametrization of
the vector $\vect{M}$: $\vect{M} = \left(\sin\theta\cos\phi,
\sin\theta\sin\phi, \cos\theta \right)$. Thus we need to introduce
a two component bosonic fields $z^{m} = (z^m_1, z^m_2)$ to
describe the hedgehog monopole of $\vect{N}$: via $\vect{M} =
\frac{1}{2}z^{m \dagger} \vect{\sigma} z^m$.

Now let us look at vector the $\vect{N} \sim (n_1, n_2, n_3)$ and
introduce the standard CP$^1$ field parametrization of $\vect{N}$:
\beqn \vect{N} = \frac{1}{2} z^{e \dagger} \vect{\sigma} z^e.
\eeqn As usual, a U(1) gauge field $a_\mu$ is introduced in this
parametrization, and when the CP$^1$ field $z^e$ is gapped and
disordered, the U(1) gauge field is in its deconfined photon
phase. In the deconfined photon phase, in addition to gapless photon
excitations, there are also two types of basic gapped point particles.
The first type is the electric charge $e$, which is the CP$^1$
field $z^e$; the second kind of gapped particle is the Dirac
monopole $m$, which is nothing but the hedgehog monopole of
$\vect{N}$. This is because in the standard CP$^1$ formalism, the
U(1) gauge flux quantum $\int d^2x \frac{1}{2\pi} \nabla \times a$
through any closed surface is just the Skyrmion number $\int d^2x
\frac{1}{4\pi} \vect{N}\cdot (\partial_x \vect{N} \times
\partial_y \vect{N}) $. Therefore the Dirac monopole, which is the
source of the gauge flux, is identified with the hedgehog
monopole, which is the source of the Skyrmion number. Thus the
Dirac monopole $m$ of the photon phase can also be represented by
$z^m$.


Now we can also turn on the symmetry $\mathcal{T}$. Based on the
transformation in Eq.~\ref{sym}, $\mathcal{T}$ interchanges the
electric charge and Dirac monopole. Under the $Z_2$ and
$\mathcal{T}$ symmetries, the electric charge $z^e$ and magnetic
monopole $z^m$ transform as \beqn Z_2 &:& z^{e,m} \rightarrow
i\sigma^y \left( z^{e,m} \right)^\ast, \cr \cr \mathcal{T}  &:&
z^{e} \rightarrow z^{m}, \ \ \ z^{m} \rightarrow z^{e}. \eeqn The
electric and magnetic field will transform as \beqn Z_2 &:&
\vect{E}, \ \vect{B} \rightarrow - \vect{E}, \ - \vect{B}; \cr\cr
\mathcal{T} &:& \vect{E} \rightarrow \vect{B}, \ \ \ \vect{B}
\rightarrow \vect{E}, \eeqn The commutation relation
$[E_i(\vect{x}), B_j(\vect{x}^\prime)] = i \epsilon_{ijk}
\partial_{x^\prime_k} \delta^3(\vect{x} - \vect{x}^\prime)$, and
hence the Maxwell equation, are invariant under the $Z_2$ and
$\mathcal{T}$ symmetries. Thus in this photon phase the
$\mathcal{T}$ symmetry acts as a electric-magnetic duality. This
photon phase cannot be driven into a trivial confined or Higgs
phase because the condensate of either $z^e$ or $z^m$ will
inevitably generate an order of a certain component of $\vect{n}$,
which therefore breaks the $Z_2$ and $\mathcal{T}$ symmetry. Since
$\mathcal{T}$ plays the role as the duality symmetry, our photon
phase is different from all the $3d$ time-reversal symmetry
enriched photon phases classified in
Ref.~\onlinecite{wangsenthil}.

The two extra SO(3) symmetries make the physical meaning of the
$e$ and $m$ particles transparent: the $e$ ($m$) particle is the
hedgehog monopole of $- \vect{M}$ ($\vect{N}$ \footnote{This
opposite sign is due to the fact that in the WZW term of
Eq.~\ref{o6wzw}, the monopole of $\vect{N}$ is a $(0+1)d$ O(3) WZW
model of $\vect{M}$ with level $k = 1$, while the monopole of
$\vect{M}$ is a WZW model of $\vect{N}$ with level $k = -1$.}),
which is also a fractionalized CP$^1$ field of $\vect{N}$
($\vect{M}$). Under the duality $Z^T_2$ symmetry, the role of $e$
and $m$ interchanges. But the nature of this photon phase does not
depend on the extra SO(3) symmetries. Thus after we establish this
photon phase, the SO(3) symmetries can be explicitly broken
without changing the physics of the photon phase.

Another point excitation of this photon phase is the dyon. A dyon
is a bound state between $e$ and $m$ which forms a fermion
(denoted as $f$). $e$ and $m$ view each other as a $2\pi-$flux
source. Therefore the effective theory that describes the internal
degree of freedom of a dyon is precisely the $(0+1)d$ O(3) NLSM
with a WZW term at level-1, $i.e.$ Eq.~\ref{0dwzw}, except with
$\vect{M}$ replaced by $\vect{D}$ where $\vect{D}$ the vector that
connects electric and magnetic charges. The ground state of this
model is again a spin-1/2 doublet. Since now $e$ and $m$
interchange under $\mathcal{T}$, this means that in this model
$\mathcal{T}$ takes $\vect{D}$ to $ - \vect{D}$, which implies
that under $\mathcal{T}$, the dyon is a Kramers doublet fermion:
$\mathcal{T}: f \rightarrow i\sigma^y f$, and $\mathcal{T}^2 =
-1$. This is due to the fact that $\mathcal{T}^2$ is effectively a
$2\pi-$rotation in space, and a spin-1/2 object acquires a minus
sign under $2\pi-$rotation. Thus the dyon transforms exactly the
same under time-reversal as a physical electron.

In the bulk of the $4d$ BSPT, the hedgehog monopole of $\vect{N}$,
or equivalently the Dirac monopole of the gauge field $a_\mu$ in
the CP$^1$ formalism, is a $1d$ loop defect. And Eq.~\ref{o6}
reduces to a $(1+1)d$ O(3) NLSM of $\vect{M}$ with a topological
$\Theta-$term with $\Theta = 2\pi$, which is an effective theory
for a spin-1 antiferromagnet chain~\cite{haldane1,haldane2}. Thus
a closed loop of hedgehog (Dirac) monopole in the $4d$ bulk is
fully gapped and nondegenerate, and can therefore proliferate and
drive the bulk into a gapped disordered SPT and gauge confined
phase. But when a monopole line terminates at the boundary, its
end cannot condense without breaking symmetry. Thus it is possible
for a deconfined photon phase to exists at the $3d$ boundary. The
physical meaning of this boundary photon phase, including its
emergent photon excitations, can also be understood in a different
way, presented in the next section.


\section{3. $\mathbb{Z}_2$ topological order with a Kramers doublet
fermion}

A $3d$ U(1) photon phase can usually be driven into a fully gapped
$\mathbb{Z}_2$ topological order by condensing either a pair of
$e$ or $m$ particles. In the condensate of $2e$, $m$ and $f$ are
confined because they have nontrivial mutual statistics with $2e$,
and the only deconfined point particle is $e$. Besides the point
particle, the system also has a gapped loop excitation (usually
called the ``vison" loop excitation), which has a mutual semionic
statistics with $e$; $i.e.$ when the closed trajectory of $e$
links with the vison loop by odd numbers, the system wave function
will acquire a minus sign. However, the vison loop has trivial
statistics with $2e$, and is therefore not confined in the $2e$
condensate.

In our system, the condensate of either $2e$ or $2m$ will break
the $\mathcal{T}$ symmetry. In order to construct a topological
order that preserves $\mathcal{T}$, we need to condense a Cooper
pair of dyons (each dyon is a fermion). Since the dyon is a
Kramers doublet, we will condense the singlet Cooper pair of the
dyons, just to preserve all the symmetries. In this condensate,
both $e$ and $m$ excitations of the original photon phase will be
confined because they have nontrivial mutual statistics with the
dyon Cooper pair (this is also called oblique
confinement~\cite{cardyoblique1,cardyoblique2}), and the only
deconfined but gapped point excitation is the fermionic dyon $f$.

The condensate of a pair of dyons also has a loop excitation which
has mutual semionic statistics with the dyon. To understand this
loop defect, let us we first turn on two extra SO(2) symmetries
where $(n_1, n_2)$ and $(n_4, n_5)$ transform as vectors under the
first and second SO(2) group respectively: \beqn \mathrm{SO(2)}_1
&:& (n_1 + in_2) \rightarrow e^{i\theta} (n_1 + in_2), \ \ z^e
\rightarrow e^{\frac{i\theta}{2} \sigma^z} z^e, \cr\cr
\mathrm{SO(2)}_2&:& (n_4 + in_5) \rightarrow e^{i\theta} (n_4 +
in_5), \ \ z^m \rightarrow e^{\frac{i\theta}{2} \sigma^z} z^m,
\eeqn This means $e$ carries half charge of the first SO(2)
symmetry and therefore must have a mutual semionic statistics with
the vortex loop of $(n_1, n_2)$; likewise, $m$ must have a mutual
semionic statistics with the vortex loop of $(n_3, n_4)$. The
dyon, which is a bound state of $e$ and $m$, must have mutual
semionic statistics with both types of vortex loops.


Besides their mutual statistics with point excitations, these
vortex loops also have a nontrivial spectrum. The $(3+1)d$ WZW term
in Eq.~\ref{o6wzw} will decorate each vortex line of $(n_1, n_2)$
with a $(1+1)d$ O(4) WZW-term at level-1: \beqn \mathcal{S}_{v} &=&
\int dx d\tau \ \frac{1}{g_2} \mathrm{tr}\left(
\partial_\mu U^\dagger \partial_\mu U \right) \cr\cr &+& \int_0^1 du \
\frac{i2\pi}{24\pi^2} \epsilon_{\mu\nu\rho} \mathrm{tr} \left(
U^\dagger \partial_\mu U U^\dagger \partial_\nu U U^\dagger
\partial_\rho U \right), \label{1dwzw} \eeqn where $U \sim n_3
\sigma^0 + i n_4 \sigma^1 + i n_5 \sigma^2 + i n_6 \sigma^3 $ is a
SU(2) matrix. The condensate of $(n_1, n_2)$ will break part of
the symmetries, but it preserves another new $Z_2^\prime $
symmetry which is the combination of a $\pi-$rotation of SO(2)$_1$
and the $Z_2$ symmetry. Then this $Z_2^\prime$ symmetry guarantees
that the vortex loop of $(n_1, n_2)$ must be either gapless or
degenerate, due to the $1d$ WZW term in Eq.~\ref{1dwzw}. The
nature of the vortex loop can also be understood as the following:
In the $4d$ bulk, a vortex of $(n_1, n_2)$ is a $2d$ membrane, and
according to the bulk action Eq.~\ref{o6}, this $2d$ membrane is
decorated with a $2d$ SPT phase with $Z_2^\prime$ symmetry, which
implies that when the vortex membrane terminates at the $3d$
boundary, it becomes a vortex loop, and it is also the boundary of
a $2d$ SPT state; thus it must be either gapless or degenerate.
The degeneracy of the vortex line corresponds to spontaneous
symmetry breaking of the $Z_2^\prime$ symmetry: for instance
$\langle n_3 \rangle > 0$ or $\langle n_3 \rangle < 0$ along the
vortex loop. Then there are two flavors of vortex loops, and the
domain wall between the two flavors is precisely the hedgehog
monopole of $\vect{N} \sim (n_1, n_2, n_3)$.

Although a single vortex loop of $(n_1, n_2)$ must have nontrivial
spectrum, a double strength vortex loop (a vortex loop of $4\pi$
vorticity) can be gapped and nondegenerate. One way to see this is
that, because the $2d$ SPT phase with $Z_2^\prime$ symmetry has a
$Z_2$ classification~\cite{levingu,wenspt,wenspt2}, two copies of
such a state becomes trivial; $i.e.$ their boundary can be rendered
gapped and nondegenerate.

Many interesting phases can be obtained by manipulating the
dynamics of the vortex loops. For example, consider a superfluid
phase with spontaneous SO(2)$_1$ symmetry breaking, $i.e.$ a
superfluid phase with condensation of complex boson $b_1 \sim
n_1+in_2$ (this phase also spontaneously breaks $\mathcal{T}$).
Then according to Ref.~\onlinecite{senthilloop}, if the vortex
loop has two flavors and both flavors proliferate, then this phase
is precisely the photon phase described in section {\bf 2}. The
$\mathbb{Z}_2$ topological order with $2e$ condensate discussed at
the beginning of this section can be realized when the strength-2
(fully gapped and nondegenerate) vortex loop of $(n_1, n_2)$
proliferates.

Now let us start with a superfluid phase where the bound state of
bosons $b_1 \sim n_1 + in_2$ and $b_2^\ast \sim n_4 - in_5$
condense. Under $\mathcal{T}$, $b_1 \rightarrow b_2$ and $b_2
\rightarrow b_1$; thus the condensate of bound state $\langle b_1
b_2^\ast \rangle$ does not break $\mathcal{T}$ ($i.e.$ the real
and imaginary parts of $\langle b_1 b_2^\ast \rangle$ are both
invariant under $\mathcal{T}$). In this phase there are two types
of vortex loops: vortex loops of $(n_1, n_2)$ and $(n_4, n_5)$.
Now we argue that the $\mathbb{Z}_2$ topological order with $2f$
condensate can be constructed by proliferating the bound state of
these two types of vortex loops. First of all, since $e$ and $m$
particles both have mutual semionic statistics with this ``bound
vortex loop", they will both be confined in this condensate; the
only deconfined point particle is the dyon $f$, which views this
``bound vortex loop" as a $2\pi-$flux instead of a $\pi-$flux
loop. Since in the $\mathbb{Z}_2$ topological order the bound
state of the two vortex loops already proliferate, there is only
one type of well-defined loop excitation, and it can be viewed as
the remnant of either the $(n_1, n_2)$ or $(n_4, n_5)$ vortex
loop, either of which has the correct semionic statistics with the
dyon. Thus this vortex loop can be identified as the vison loop
excitation of the desired $3d$ $\mathbb{Z}_2$ topological order.

More systematically, the condensate of bound state $b_1 b^\ast_2$
can be described by the following the effective action in the
Euclidean space-time: \beqn \mathcal{S} = \sum_{\vect{x}}
\sum_{c=1,2} - K\cos(\dd \theta_c-a), \eeqn where $b_{1,2} \sim
\exp(i\theta_{1,2})$, and $a$ is a 1-form gauge field. In this
equation, when both $b_1$ and $b_2$ condense, the only gauge
invariant order parameter is $b_1 b_2^\ast$.

We can take the standard Villain form of the action by expanding
the cosine function at its minimum and introducing the 1-form
fields $l_c\in\dsZ$ and $k_c\in\mathbb{R}$ ($c=1,2$): \beq
\begin{split}
\mathcal{Z}=&\Tr \exp\Big[\sum_{\vect{x}} \sum_{c=1,2} -\frac{K}{2}(\dd \theta_c-a-2\pi l_c)^2\Big]\\
\sim&\Tr \exp\Big[\sum_{\vect{x}} \sum_{c=1,2} -\frac{1}{2K}k_c^2+i k_c\cdot(\dd \theta_c-a-2\pi l_c)\Big]\\
\sim&\Tr \exp\Big[\sum_{\vect{x}} \sum_{c=1,2} -\frac{1}{2K}k_c^2-i k_c\cdot(a+2\pi l_c)\Big]\delta[\partial k_c]\\
\sim&\Tr \exp\Big[\sum_{\vect{x}} \sum_{c=1,2} -\frac{1}{2K}(\dd
\Omega_c)^2+i (a+2\pi l_c)\wedge \dd \Omega_c\Big].
\end{split}
\eeq In the last line, we introduce the 2-form fields $\Omega_c$
($c=1,2$) on the dual space-time manifold, such that $k_c=\star\dd
\Omega_c$ resolves the constraint $\partial k_c=0$. Summing over
$l_c$ will require $\Omega_c$ to take only integer values, which
could be imposed by adding a $\cos(2\pi \Omega_c)$ term, and the
theory now becomes \beq \mathcal{Z}\sim\Tr
\exp\Big[\sum_{\vect{x}} \sum_{c=1,2} -\frac{1}{2K}(\dd
\Omega_c)^2+i a\wedge \dd \Omega_c+t\cos(2\pi \Omega_c) \Big] \eeq
Integrating out the gauge field $a$ will impose the constraint
$\dd(\Omega_1 + \Omega_2)=0$, which can be resolved by $\Omega_1 =
\Omega - \dd v_1/(2\pi)$, $\Omega_2 = - \Omega + \dd v_2/(2\pi)$.
Therefore the final action takes the form of \beqn \mathcal{S}
\sim \sum_{\vect{x}} \sum_{c = 1,2} - t \cos(\dd v_c - 2\pi
\Omega) + \frac{1}{K}(\dd \Omega)^2. \eeqn

$v_1$ and $v_2$ are both 1-form vector fields. $\Psi_{1,\mu} \sim
\exp(iv_{1,\mu})$ creates a segment of vortex line of $b_1$ along the
$\hat{\mu}$ direction, while $\exp(i \dd v_1)$ creates a unit
vortex loop. If we are going to proliferate the bound state of the
two types of vortex loops, say $\Psi_{1,\mu}\Psi_{2,\mu} \sim
\exp(iv_\mu) = \exp(i v_{1,\mu} + i v_{2,\mu})$, then the
effective action for $v_\mu$ reads \beqn \mathcal{S} \sim
\sum_{\vect{x}} - t \cos(\dd v - 4\pi \Omega) + \frac{1}{K}(\dd
\Omega)^2. \eeqn When $v_\mu$ proliferates, $\Omega$ can take two
inequivalent minima: $\Omega = 0, 1/2$; therefore this state is a
$\mathbb{Z}_2$ topological order.

Once we establish the existence of this $\mathbb{Z}_2$ topological
order, the extra SO(2) symmetries can be broken, which will not
affect the statistics between vison loops and the dyon $f$. In
this $\mathbb{Z}_2$ topological order, since there is no
spontaneous symmetry breaking at all, the original $Z_2$ symmetry
already guarantees that the vison loop must be either gapless or
degenerate because the vison loop is effectively the boundary of a
$2d$ SPT state with $Z_2$ symmetry. However, after we break the
two SO(2) symmetries, the bound state between the $(n_1, n_2)$ and
$(n_4, n_5)$ vortex loops become gapped and nondegenerate,
allowing it to safely proliferate. Because the deconfined point
particle excitation of this $\mathbb{Z}_2$ topological order is a
fermion, it cannot condense and drive the system into a trivial
Higgs phase; similarly, because the vison loop is gapless or
degenerate, it also cannot proliferate and drive the system into a
gapped and nondegenerate confined phase.

While the $\mathbb{Z}_2$ topological order itself cannot be driven
into either a trivial Higgs or confined phase, two copies of this
$\mathbb{Z}_2$ topological order can indeed be trivialized. The
reason is that, for two copies of the $\mathbb{Z}_2$ topological
order (labeled $A$ and $B$), one can first condense the bound
state of dyons from both copies ($i.e.$ condensate of $f_Af_B$) to
break the two copies of $\mathbb{Z}_2$ topological order down to
one $\mathbb{Z}_2$ topological order. Then in this residual
$\mathbb{Z}_2$ topological order the only well-defined point
particle is the dyon $f_A$ (or equivalently $f_B$ because of the
background pair condensate). The vison loop is the bound state of
vison loops from both copies: $\Psi_{A,\mu} \Psi_{B,\mu}$; this is
because $\Psi_{A,\mu}$ and $\Psi_{B,\mu}$ individually have
semionic statistics with $f_Af_B$, and hence must be confined in
the condensate. Since $\Psi_{A,\mu}$ and $\Psi_{B,\mu}$ both carry
a $(1+1)d$ WZW term at level-1, their bound state is fully gapped,
and hence can further proliferate and drive the entire system into
a trivial confined phase without any symmetry breaking. This
implies that two copies of the BSPT states Eq.~\ref{o6} with $Z_2
\times \mathcal{T}$ symmetry is a trivial state, which is
consistent with the classification based on the NLSM itself given
in appendix {\bf A}.

Having understood the self-dual photon phase and the
$\mathbb{Z}_2$ topological order, we can realize an exotic
$(3+1)d$ self-dual topological photon phase by making a thin slab
of $4d$ BSPT state with $Z_2 \times \mathcal{T}$ symmetry, and
realizing the self-dual photon phase on the top boundary and the
$\mathbb{Z}_2$ topological order on the bottom boundary. At low
energy, only the photon phase at the top boundary will be
detectable while only at higher energy will the bottom boundary be
exposed.


\section{4. Self-dual photon phase with $Z_4$ symmetry}

Another BSPT phase that leads to a self-dual photon phase at its
boundary is a state with $Z_4$ symmetry, which is still described
by Eq.~\ref{o6}, but now the the vector $\vect{n}$ transforms as
\beqn Z_4 &:& (n_1 + in_2) \rightarrow e^{i\theta} (n_1 + in_2),
\cr\cr && (n_3 + in_4) \rightarrow e^{i\theta} (n_3 + in_4),
\cr\cr && (n_5 + in_6) \rightarrow e^{i\theta} (n_5 + in_6), \eeqn
where $\theta = \frac{k \pi}{2}$, with $k = 0, 1, 2, 3$. Using the
formalisms introduced in section {\bf 2}, we can demonstrate that
the $3d$ boundary of this $4d$ BSPT is a self-dual photon phase
with the following transformation of its $e$ and $m$ excitations:
\beqn Z_4: z^e \rightarrow z^m, \ \ \ z^m \rightarrow i\sigma^y
(z^e)^\ast.  \eeqn Note here $e$ and $m$ are hedgehog monopoles of
three component vectors $\vect{N} \sim (n_2, n_4, n_6)$ and
$\vect{M} \sim (n_1, n_3, n_5)$ respectively. The emergent
electric and magnetic fields transform under the $Z_4$ symmetry as
\beqn \vect{E} \rightarrow \vect{B}, \ \ \ \vect{B} \rightarrow -
\vect{E}. \eeqn Again the Maxwell equation and the commutation
relation between $\vect{E}$ and $\vect{B}$ fields are invariant
under this $Z_4$ transformation.

Unlike the previous case, it is not obvious whether we can drive
this this self-dual photon phase into a gapped topological order
with full $Z_4$ symmetry. As we discuss in appendix {\bf B}, this
$Z_4$ BSPT state can be constructed by breaking the U(1) symmetry
of the $4d$ bosonic integer quantum Hall state down to $Z_4$. In
Ref.~\onlinecite{xuglobal}, we argued that a BSPT state whose
boundary has perturbative gauge anomaly after ``gauging" the
symmetry cannot be driven into a fully gapped topological order
because the system must respond to infinitesimal external gauge
field. The boundary of the $4d$ bosonic integer quantum Hall
(BIQH) state we discuss in appendix {\bf B} has a perturbative
gauge anomaly after the U(1) symmetry is gauged. Thus the boundary
of the $4d$ BIQH state cannot be driven into a symmetric
topological order. It is possible that the $Z_4$ BSPT state
inherits this property from its BIQH parent state. More rigorous
study will be given in the future.

\section{5. Summary}

In this work we studied two examples of different self-dual photon
phases that can be realized on the boundary of $4d$ bosonic SPT
states. Both states need certain symmetries to protect their
boundaries, which is an important difference from the self-dual
photon phase studied in Ref.~\onlinecite{mcgreevy2014}.
Understanding these symmetry protected self-dual photon phases at
the boundary of $4d$ systems can lead to exotic photon phases on a
$3d$ system as well, as was discussed in the end of section {\bf
3}. In even higher dimensions, topological orders and photon
states can still exist on the boundary of BSPT states; but more
exotic boundary states can also be realized, such as deconfined
spin liquid phases with nonabelian gauge fields. This is due to
the fact that a nonabelian gauge fields usually lead to
confinement in dimensions lower than $3+1$, while in higher
dimensions a stable deconfined phase exists.

The authors are supported by the the David and Lucile Packard
Foundation and NSF Grant No. DMR-1151208.


\bibliography{QED}

\section{Supplemental Material}
\maketitle
\appendix

\section{A. NLSM field theory description of BSTP states}

Bosonic SPT phases in {\it all dimensions} with various continuous
and discrete symmetries can be systematically described and
classified by semiclassical nonlinear Sigma model (NLSM) field
theories with a topological $\Theta$-term~\cite{xuclass}.

In $1+1d$, it is well known that a spin-1 chain can be described by an
$O(3)$ NLSM with topological a $\Theta$-term
\cite{haldane1,haldane2,affleck1987,kennedy1990,hagiwara1990,ng1994}
at $\Theta=2\pi$ where its disordered phase corresponds to a BSPT
state protected by $O(3)$ or time reversal symmetry. In general, a
BSPT phase in $(d+1)-$dimensional space-time can always be
formulated by an O($d+2$) NLSM with a topological $\Theta$-term,
assuming the symmetry group $G$ of the BSPT state is a subgroup of
$O(d+2)$ and other discrete symmetries such as time reversal:
\begin{align} \mathcal{S}^{\Theta}_{d+1} &= \int d^dx d\tau \ \frac{1}{g}
(\partial_\mu \vect{n})^2 \\ &+ \frac{i\Theta}{\Omega_{d+1}}
\epsilon_{i_1,...,i_{d+2}} n^{i_1} \partial_{x_1} n^{i_2} \cdots
\partial_{x_d} n^{i_{d+1}}\partial_{\tau} n^{i_d+2} \nonumber
\label{ontheta} \end{align}

The boundary theory of the $d+1$-dimensional theories with
$\Theta=2\pi$ are described by $(d-1)+1-$dimensional O($d+2$)
NLSMs with a Wess-Zumino-Witten (WZW) term at level-1:
\begin{align} \mathcal{S}^{WZW}_{d+1} &= \int d^{d-1}x d\tau \
\frac{1}{g} (\partial_\mu \vect{n})^2 \\
&+\int_0^1du\frac{i2\pi}{\Omega_{d+1}} \epsilon_{i_1,...,i_{d+2}}
n^{i_1} \partial_{x_1} n^{i_2} \cdots \partial_{\tau}
n^{i_{d+1}}\partial_{u} n^{i_d+2} \nonumber \label{onwzw}
\end{align} To define the WZW-term we need to extend the order
parameter field $\vect{n}(x_1,x_2,...,x_{d-1},\tau)$ to
$\vect{n}(x_1,x_2,...,x_{d-1},\tau,u)$ with the following
condition: \beqn
\vect{n}(x_1,x_2,...,x_{d-1},\tau,0)&=&(0,0,...,1)\\
\vect{n}(x_1,x_2,...,x_{d-1},\tau,1)&=&\vect{n}(x_1,x_2,...,x_{d-1},\tau)
\eeqn

The spectrum of the boundary theory above is in general
non-trivial: either degenerate or gapless, provided we have enough
symmetry. For example, when $d = 1$, the boundary is a 0+1d O(3)
NLSM with a WZW term at level-1. If the theory has full $O(3)$
symmetry or time reversal symmetry, the ground state is a doublet
with protected two-fold degeneracy. When $d = 2$, the boundary is
a 1+1d $O(4)$ WZW-term at level-1, which is conformal assuming the
full $O(4)$ symmetry is preserved
~\cite{witten1984,KnizhnikZamolodchikov1984}. The spectrum can
also be degenerate if we only have discrete symmetry and the
degeneracy is precisely due to the symmetry breaking.

Notice that all components of $\vect{n}$ in Eq.~\ref{ontheta} must
have a nontrivial transformation under the symmetry group $G$.
Otherwise one can turn on a linear ``Zeeman" term that polarizes
some component of $\vect{n}$ which will trivially gap out the edge
states. In this case, the $\Theta$-term has no effect, and the
bulk state is trivial.

Eq.~\ref{o6} with $Z_2 \times \mathcal{T}$ symmetry is a
nontrivial BSTP state when $\Theta = 2\pi$. However, two copies
(layers) of Eq.~\ref{o6} can be trivialized after turning on
symmetry allowed interlayer couplings. For instance, starting with
two copies of Eq.~\ref{o6} (labeled $A$ and $B$), the following
coupling is allowed by the symmetry: \beqn H_{AB} &=& \int d^4x \
- w (n_{A,1} n_{B,4} + n_{A,2} n_{B,5} + n_{A,3} n_{B,6} \cr\cr
&+&  n_{A,4} n_{B,1} + n_{A,5} n_{B,2} + n_{A,6} n_{B,3} ).
\label{abw}\eeqn When $w$ is positive and large, \beqn (n_{A, 1},
n_{A, 2}, n_{A, 3}) \sim (n_{B, 4}, n_{B, 5}, n_{B, 6}), \cr\cr
(n_{A, 4}, n_{A, 5}, n_{A, 6}) \sim (n_{B, 1}, n_{B, 2}, n_{B,
3}), \eeqn As a result, the two $\Theta-$terms of copies $A$ and
$B$ will cancel out, and effectively the coupled system has
$\Theta = 0$, and is therefore a trivial state. This conclusion is
consistent with the analysis based on the boundary topological
orders in section {\bf 3}.

\section{B. $4d$ bosonic integer quantum Hall state as parent state}

In this section we discuss $4d$ bosonic integer quantum Hall
(BIQH) states and their relation with the two states discussed in
this work. The $4d$ BIQH state is a straightforward generalization
of the $2d$ BIQH state discussed in
Ref.~\onlinecite{levinsenthil}. It is described by a $(4+1)d$ O(6)
NLSM with $\Theta = 2\pi$ (Eq.~\ref{o6}), where the six component
vector $\vect{n}$ transforms under the U(1) symmetry as \beqn U(1)
&:& (n_1 + in_2) \rightarrow e^{i\theta} (n_1 + in_2), \cr\cr &&
(n_3 + in_4) \rightarrow e^{i\theta} (n_3 + in_4), \cr\cr && (n_5
+ in_6) \rightarrow e^{i\theta} (n_5 + in_6). \eeqn If we couple
the U(1) charge to an external U(1) gauge field $A_\mu$, then
after integrating out the matter field $\vect{n}$, a Chern-Simons
term is generated for $A_\mu$: \beqn \mathcal{S}_{cs} = \int d^4x
d\tau \ \frac{6 i}{3! (2\pi)^2} \epsilon_{\mu\nu\rho\alpha\beta}
A_\mu
\partial_\nu A_{\rho} \partial_\alpha A_\beta,
\label{cs} \eeqn
which is a CS theory at level 6.

Directly integrating out the boson field is technically difficult.
But alternatively we can start with 8 copies of $4d$ fermionic
integer quantum Hall model: \beqn H = \sum_{\vect{k}} \sum_{a =
1}^8 \psi^\dagger_{a,\vect{k}} \left( \sum_{i = 1}^4 \Gamma_i \sin
k_i +  m ( e - \sum_{i} \cos k_i ) \Gamma_5 \right)
\psi_{a,\vect{k}} \label{4d}\eeqn $\Gamma_1 = \sigma^{13}$,
$\Gamma_{2} = \sigma^{23}$, $\Gamma_3 = \sigma^{33}$, $\Gamma_4 =
\sigma^{01}$, $\Gamma_5 = \sigma^{02}$ where $\sigma^{ab\cdots} =
\sigma^a \otimes \sigma^b \otimes \cdots$. We focus on the phase
with $ 3 < e < 4 $, where each fermion copy (labeled by $a$) gives
rise to a $3d$ chiral fermion at the boundary. Therefore there are
in total 8 chiral fermions at the boundary of Eq.~\ref{4d}: \beqn
H_{3d} = \int d^3x \ \sum_{a = 1}^8 \psi^\dagger_a ( i
\vect{\sigma} \cdot \vect{\partial}) \psi_a. \eeqn

Now we can couple the boundary chiral fermions to a six component
vector $\vect{n}$:
\begin{align} H_{int} =& \int d^3x \nonumber\\
& u (n_1 \mathrm{Re}[\psi^t \sigma_2 \otimes \sigma^{122} \psi] + n_2
  \mathrm{Im}[\psi^t \sigma_2 \otimes \sigma^{212} \psi]) \nonumber\\
+& u ( n_3 \mathrm{Re}[\psi^t \sigma_2 \otimes \sigma^{202} \psi] +
  n_4 \mathrm{Im}[\psi^t \sigma_2 \otimes \sigma^{022} \psi]) \nonumber\\
+& u ( n_5 \mathrm{Re}[\psi^t \sigma_2 \otimes \sigma^{322} \psi]
  + n_6 \mathrm{Im}[\psi^t \sigma_2 \otimes \sigma^{232} \psi]).
\end{align}
The same WZW term as Eq.~\ref{o6wzw} will be generated after
integrating out the fermions.

This fermion model (Eq.~\ref{4d}) has at most a U(8) symmetry,
which contains three U(1) symmetries as a subgroup. The three U(1)
symmetries are generated by $\sigma^{330}$, $\sigma^{220}$, and
$\sigma^{110}$ so that $(n_1, n_2)$, $(n_3, n_4)$, and $(n_5,
n_6)$ transform as two-component vectors under these three U(1)
symmetries, respectively. Now let us couple the fermion model
Eq.~\ref{4d} to three U(1) gauge fields: $\frac{1}{2} A^{(1)}_\mu
\sigma^{330}$, $\frac{1}{2} A^{(2)}_\mu \sigma^{220}$,
$\frac{1}{2} A^{(3)}_\mu \sigma^{110}$. We give the fermions
charge$-1/2$ because we want the bosons to carry charge$-1$ under
these gauge fields. Then after integrating out the fermions, the
following Chern-Simons field theory is generated: \beqn
\mathcal{S} &=& \frac{1}{8}
\mathrm{tr}[\sigma^{330}\sigma^{220}\sigma^{110}] \int d^4x d\tau
\cr\cr &\times& \frac{i}{3! (2\pi)^2}
\epsilon_{\mu\nu\rho\alpha\beta} A^{(1)}_\mu \partial_\nu
A^{(2)}_{\rho} \partial_\alpha A^{(3)}_\beta \cr\cr &+&
\mathrm{permutation} \ \mathrm{of} \ 1,2,3; \eeqn After breaking
these three U(1) gauge symmetries down to a single U(1) gauge
symmetry, Eq.~\ref{cs} is generated.

The two BSPT states we discussed in this paper can be obtained by
breaking the U(1) symmetry down to either $Z_2$ or $Z_4$ symmetry.
Notice that the BIQH state with U(1) symmetry has a $\mathbb{Z}$
classification with U(1) symmetry and $\Theta = 2\pi k$ where
each integer $k$ corresponds to a different BIQH state. Notice
that the coupling Eq.~\ref{abw} explicitly breaks the U(1)
symmetry.

\section{C. Dimensional Reduction of Topological Terms}

In this Appendix, we are going to derive the effective field
theory of a monopole core of an O(6) NLSM, namely Eq.~\ref{0dwzw}.
In $3+1d$ a monopole configuration of an O(3) order parameter, for
instance $(n_1,n_2,n_3)$, can be understood as an intersection
point of the domain walls of the three order parameter fields
respectively. So we can derive the theory on a monopole core by
three domain wall projections, which is described below.

To derive the theory on the domain wall of one of the order
parameter fields, e.g. $n_1$, we'll first construct a domain wall
configuration of $n_1$. Consider the following configuration of
the vector $\vect{n}$: $\vect{n} = (\cos\theta, \sin\theta N_2,
\sin\theta N_3, \sin\theta N_4, \sin\theta N_5,\sin\theta N_6)$,
where $\vect{N}$ is an O(5) vector with unit length and $\theta$
is a function of coordinate $z$ only with \beqn \theta(z =
+\infty) = \pi, \ \ \ \theta(z = - \infty) = 0. \eeqn By inserting
this parametrization of $\vect{n}$ into Eq.~\ref{o6wzw} and
integrating along the $z$ direction, the $O(6)$ WZW-term reduces
to an $O(5)$ WZW-term with the same level. More explicitly, the
theory on the domain wall is: \beqn \mathcal{S}_{dw} &=& \int d^2x
d\tau \ \frac{1}{g} (\partial_\mu \vect{N})^2 \cr &+& \int_0^1 du
\frac{i2\pi}{\Omega_4} \epsilon_{abcde} N^a \partial_{x} N^b
\partial_{y} N^c \partial_{\tau} N^d \partial_{u} N^e . \label{o5wzw}
\eeqn A domain wall projection reduces both the spatial dimension
and the dimension of the order parameter field by one, and the
effective field theory on the domain wall inherits the topological
term from the original theory.

We can repeat this domain wall projection procedure once more. On
the $n_1$ domain wall we just made, consider a domain wall of
$n_2$ along the $y$-direction. We can integrate over the
$y$-direction, and the resulting theory is an $O(4)$ WZW-term with
level-1: \beqn \mathcal{S}_{v} &=& \int d^1x d\tau \ \frac{1}{g}
(\partial_\mu \vect{N})^2 \cr &+& \int_0^1 du
\frac{i2\pi}{\Omega_3} \epsilon_{abcd} N^a \partial_{x} N^b
\partial_{\tau} N^c \partial_{u} N^d . \label{o4wzw}
\eeqn This field theory can be thought of as the effective field
theory on a $2\pi$-vortex of $(n_1,n_2)$ components. Notice that this
field theory is equivalent to an $1+1d$ $SU(2)$ principle chiral
model by introducing $SU(2)$ matrix field
$U=n_3\sigma^0+in_4\sigma^1+in_5\sigma^2+in_6\sigma^3$. The
$SU(2)$ principle chiral model is precisely written as in
Eq.~\ref{1dwzw}.

Based on the configuration we already have, if we further make a
domain wall of $n_3$ on the vortex core, then the whole
configuration of the order parameter field corresponds to a
monopole configuration of the $O(3)$ order parameter. And right on
the core of the monopole, the effective field theory is precisely
an $O(3)$ WZW-term at level-1 as in Eq.~\ref{0dwzw}.


\end{document}